\documentclass{emulateapj}
\usepackage{natbib}
\shorttitle{Episodic starbursts in dSphs}
\shortauthors{Nichols, Lin \& Bland-Hawthorn}

\newcommand{\HI}{\ifmmode{{\rm H\scriptstyle I}}\else{H${\scriptstyle\rm I}$}\fi}
\newcommand{\HIsub}{\rm{H}\scriptscriptstyle{I}}
\def\lta{\;\lower 0.5ex\hbox{$\buildrel < \over \sim\ $}}
\begin{document}

\title{Episodic starbursts in dwarf spheroidal galaxies: a simple model}

\author{Matthew Nichols}
\email{m.nichols@physics.usyd.edu.au}
\affil{Sydney Institute for Astronomy, School of Physics, The University of Sydney, NSW 2006, Australia\vspace*{-0.2in}}
\author{Doug Lin}
\affil{Theoretical Astrophysics Santa Cruz (TASC), Department of Astronomy and Astrophysics, University of California, Santa Cruz, CA 95064, USA}
\and
\author{Joss Bland-Hawthorn}
\affil{Sydney Institute for Astronomy, School of Physics, The University of Sydney, NSW 2006, Australia}

\begin{abstract}
Dwarf galaxies in the Local Group appear to be stripped of their gas within $270$~kpc of the host galaxy.
Color-magnitude diagrams of these dwarfs, however, show clear evidence of episodic star formation ($\Delta{}t\sim{}$a few Gyr) over cosmic time.
We present a simple model to account for this behavior.
Residual gas within the weak gravity field of the dwarf experiences dramatic variations in the gas cooling time around the eccentric orbit.
This variation is due to two main effects.
The azimuthal compression along the orbit leads to an increase in the gas cooling rate of $\sim$$([1+\epsilon]/[1-\epsilon])^2$.
The Galaxy's ionizing field declines as $1/R^2$ for $R>R_{\rm disk}$ although this reaches a floor at $R\sim150$~kpc due to the extragalactic UV field ionizing intensity.
We predict that episodic SF is mostly characteristic of dwarfs on moderately eccentric orbits $(\epsilon>0.2)$ that do not come too close to the center $(R>R_{\rm disk})$ and do not spend their entire orbit far away from the center $(R\ga200$~kpc$)$.
Up to $40\%$ of early infall dwarf spheroidals can be expected to have already had at least one burst since the initial epoch of star formation, and $10\%$ of these dwarf spheriodals experiencing a second burst.
Such a model can explain the timing of bursts in the Carina dwarf spheroidal and restrict the orbit of the Fornax dwarf spheroidal.
However, this model fails to explain why some dwarfs, such as Ursa Minor, experience no burst post-infall.
\end{abstract}

\section{Introduction}

Dwarf galaxies, being some of the earliest collapsed structures in $\Lambda$CDM and hence representing potential building blocks of larger galaxies, have come to the forefront of studies of galactic evolution in recent years.
The more than doubling of known Local Group dwarfs since the release of SDSS \citep{York2000} has provided many more dwarfs in varying environmental situations to study.
Despite most dwarfs being gas deficient within $270$~kpc of the Galaxy \citep{Grcevich2009} they all show signs of ancient star formation potentially explaining this depletion of gas \citep{Nichols2011}.
However, in addition to these ancient stellar populations many of these dwarfs around the Galaxy and within the Local Group show signs of distinct populations of younger stars arising from multiple starbursts separated by gigayears \citep{Tolstoy2009,Weisz2011}.
For example, Fornax dwarf galaxy contains stars younger than $1$~Gyr \citep{Coleman2008}.
Paradoxically, there is little, if any, trace of molecular or atomic hydrogen to provide a sufficient reservoir of cool gas to enable the onset of star burst activities \citep[Fornax shows only an off center cloud which may be Galactic gas, Carina with similar bursts shows no gas;][]{Grcevich2009}.

These bursts of star formation are together responsible for, on average, $25\%$ of the star formation within a dwarf \citep{Lee2009} and may last for a period of time of order the dynamical time of the system \citep{McQuinn2010b}.
The length and timing of these bursts have a large variation between dwarfs with some only experiencing one early burst of star formation, to Carina-like dwarfs with several distinct periods of star formation \citep{Dolphin2005}, to those that have continuous star formation with several to no small bursts \citep{Weisz2011}.

Stars formed over several generations also appear to have diverse heavy element abundance.
The stellar metallicity distribution in many dwarfs is consistent with that expected from self contamination by early generations of massive stars, albeit with a substantial loss of supernova ejecta \citep{Kirby2011,Kirby2011b}.
In isolated dwarfs far from a massive spiral or elliptical galaxy, these bursts may be a consequence of gas being blown out and subsequently infalling with a time period of several gigayears \citep{Dong2003,Valcke2008,Revaz2009}.
Episodic star formation in these isolated dwarfs require the gas that is blown out to remain bound.
Otherwise if the dwarf is orbiting a host system, the gas could easily be blown outside of the Roche sphere to fall onto the host galaxy.
Although radiatively driven outflow and tidal disruption may account for the lack of gas in these dwarf galaxies today, they also highlight the difficulties in gas retention and self contamination.

Here we present a simple self-consistent model that may explain not only star bursts in non-isolated environments but also the lack of neutral gas in the dwarf galaxies around the Galaxy.
We suggest gas is expelled after a star burst and subsequently reaccreted by its original host dwarf galaxies.
This reaccretion process only occurs at the apogalacticon and any expelled gas that is not shielded from stripping by the dwarf is lost to the host.

\section{Model}
For a dwarf galaxy to successfully accrete gas at apogalacticon, the gas must stay within the dwarfs gravitational influence for a sufficiently long time to be allowed to cool and collapse before the tidal forces and ionizing radiation field are felt as the dwarf approaches perigalacticon.

The Roche sphere is the volume of space around the dwarf within which the dwarfs gravitational pull exceeds that of the Galaxy.
At any point along the orbit the radius of the instantaneous Roche sphere is given by \citep{King1962}
\begin{equation}
  r_{\rm RS} = \left[\frac{Gm}{\omega^2-{\rm d}^2V/{\rm d}R^2}\right]^{1/3},
\end{equation}
where $r_{\rm RS}$ is the radius of the Roche sphere, $G$ the gravitational constant, $m$ the mass of the dwarf, $\omega$ the angular velocity, $V$ the potential of the Galaxy and $R$ the Galactocentric radius.

At perigalacticon and apogalacticon in a Keplerian orbit, the size of the Roche sphere simplifies to
\begin{eqnarray}
  r_{\rm RS}(R_{\rm peri}) &=& R_{\rm peri}\left[\frac{m}{M(3+\epsilon)}\right]^{1/3},\label{eq:rperi}\\
  r_{\rm RS}(R_{\rm apo}) &=& \frac{1+\epsilon}{1-\epsilon}R_{\rm peri}\left[\frac{m}{M(3-\epsilon)}\right]^{1/3}\label{eq:rapo},
\end{eqnarray}
where $R_{\rm peri}$ and $R_{\rm apo}$ is the radius of perigalacticon and apogalacticon respectively, $M$ is the mass of the galaxy and $\epsilon$ the eccentricity of the orbit.

\begin{figure}
  \begin{center}
    \includegraphics[width=0.5\textwidth]{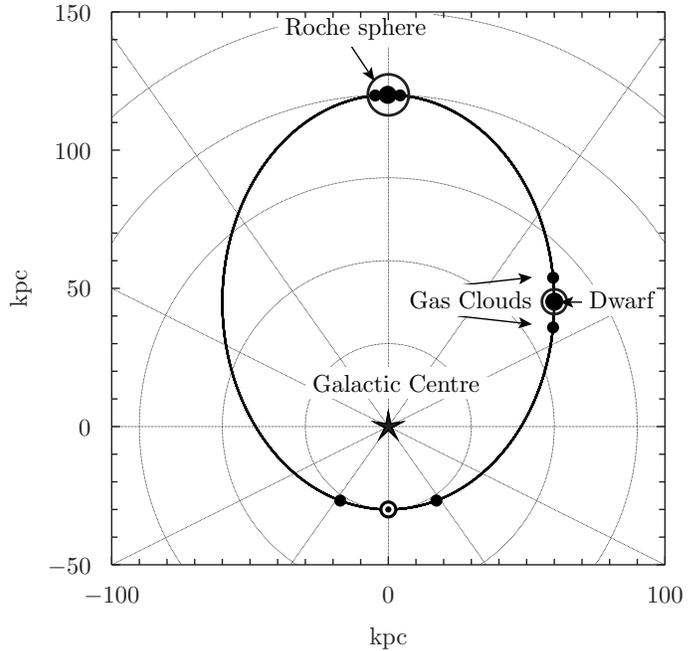}
    \caption{The change in Roche sphere over the orbit of a dwarf with a perigalacticon of $30$~kpc and eccentricity of $0.6$.
    The dwarf is represented by a solid circle in between two (smaller) solid circles representing gas clouds.
    The Roche sphere for a $3\times10^8$~$M_\odot$ dwarf around a $5.3\times10^{11}$~$M_\odot$ galaxy is shown as a circle surrounding the dwarf at apogalacticon and at the mid-way point.
    At the perigalacticon the Roche sphere is shown as a white circle inside the point representing the dwarf galaxy.
    The Roche sphere changes from being able to encapsulate points at $\pm50$~Myr along the orbit to being an order of magnitude too small at perigalacticon.
    The orbital compression is visible by the changing spacing along the orbit, a consequence of Kepler's second law.
    \label{fig:roche}}
  \end{center}

\end{figure}

This increase in the Roche sphere, shown schematically in Figure \ref{fig:roche}, allows gas that is unbound at perigalacticon to be nominally bound at apogalacticon (assuming the momentum of the gas carries it through to apogalacticon).
This gas is both compressed by the orbital path (by a factor of $[1+\epsilon]/[1-\epsilon]$) and experiences a weaker radiation field far out from the Galactic center.
These factors allow the gas to cool and fall into the potential well of the dwarf triggering a starburst.

In order to participate in reaccretion, gas that is blown out must first survive an orbit of the Galaxy.
This gas, which is likely to form small clouds once removed from the dwarf \citep{Mayer2006}, must survive the combination of high velocities and a strong radiation field at perigalacticon.
For a cloud of gas with uniform density in hydrostatic equilibrium that is located directly behind the dwarf to survive an orbit with the same velocity as it began it must have an average distance, $R_{\rm cloud}$, behind the dwarf of
\begin{equation}
  R_{\rm cloud}\sim \sqrt{\frac{2R_{\rm peri}(1+\epsilon)mr_{\rm cloud}}{C_{D}M}\frac{T_{\rm halo}}{T_{\rm cloud}}},
\end{equation}
where $r_{\rm cloud}$ is the radius of the cloud, $m$ and $M$ are the masses of the dwarf and Galaxy respectively, $T_{\rm halo}$ and $T_{\rm cloud}$ are the temperatures of the halo and cloud respectively and $C_{D}$ is the coefficient of drag of the cloud ($\sim$$1$).
Assuming that the mass ratio of the dwarf to the Galaxy is not too large, this value is comparable to the Roche sphere at perigalacticon.
However, material off center may have to be much closer to the dwarf to experience the same level of protection from drag.
The radius of survival will be increased by the dwarf's supersonic movement through the hot halo, and by a turbulent wake created by warm and hot gas that is removed from the dwarf.
These factors result in a much lower free-stream velocity \citep{Neve1989}, and subsequently lower drag, behind the dwarf for gas clouds that have been expelled by the dwarf.
As trailing gas clouds rely on both the shock generated by the dwarf and the turbulent wake created by dwarfs passage, a column of protection is formed behind the dwarf approximately the size of the Roche sphere dissipating at large distances.
Gas clouds located in front of the dwarf will be slowed by drag until they are recaptured by the dwarf or they sink into the Galactic potential well.
As gas clouds must be within the Roche sphere of the dwarf in order to be captured, a similar column of `protection' (although in this case, a column of recapture) is located ahead of the dwarf.
We therefore approximate the amount of material that is captured by the dwarf or survives an orbit as the amount of gas inside a column of radius equal to the Roche sphere at perigalacticon and extending along the orbital path of the dwarf.

Due to the vigorous star formation taking place throughout a burst we assume that all gas within the dwarf becomes ionized and is expelled from the centre.
This gas, having been heated by the burst of star formation, will expand spherically and is assumed to be constant density.
As this sphere would be confined by the pressure of the hot halo, we further assume that the density of the sphere at burst $i$ is the same as the density at the previous burst, $i-1$, i.e. $n_{{\rm H},i-1}=n_{{\rm H},i}$.
The growing hot Galactic halo and the infusion of metals into the expelled gas after each burst may alter this assumption slightly.
If the mass present in the last burst was $M_{{\rm gas},i-1}$ then the sphere of gas post burst has a radius of $r_{g,i-1}=[(3M_{{\rm gas},i-1})/(4\pi{}m_{\rm H}n_{\rm H})]^{1/3}$, where $m_{\rm H}$ is the mass of a hydrogen atom and $n_{\rm H}$ the density of hydrogen.
As the post-burst density of gas in this sphere is constant the only time this sphere can exceed the Roche sphere at apogalacticon is after the initial period of star formation, i.e. $r_{g,i}\le{}r_{\rm RS}(R_{\rm apo})$, $i>1$.

Not all gas that fills this sphere will end up being redistributed after the next starburst. 
In addition to gas lost to stripping, some gas will also be consumed in the starburst ending up as the constituent stars.
We assume the gas that ends up as stars to be approximately the mass in the Roche sphere at perigalacticon.
The gas that occupies the Roche sphere at perigalacticon may also be providing low level star formation throughout the orbit.
This mass, chosen for analytic convenience, is---for most orbits---similar to the mass that would be consumed in a typical starburst given the gas consumption timescales and burst lengths \citep{McQuinn2010,McQuinn2010b}.

The radius in the warm sphere post burst is then
\begin{eqnarray}
  \frac{4}{3}\pi{}r^3_{{\rm gas},i} &=& 2\pi{}r^2_{\rm RS}(R_{\rm peri})r_{{\rm gas},i-1} - 2\pi{}r^3_{\rm RS}(R_{\rm peri}),\nonumber \\
  r_{{\rm gas},i} &=& r_{\rm RS}(R_{\rm peri})\left[\frac{3}{2}\left(\frac{r_{{\rm gas},i-1}}{r_{\rm RS}(R_{\rm peri})} - 1\right)\right]^{1/3}, \label{eq:Eq5}
\end{eqnarray}
with the volume of the Roche sphere at perigalacticon is altered by a factor of $3/2$ to account for the difference in volume between a sphere and cylinder as $r_{{\rm gas},i}/r_{{\rm gas},i-1}\rightarrow1$.

As this gas is expelled from the dwarf by the starburst, it will quickly be heated to a warm phase of around $T=10^{4}$~K.
This gas will be moving through an isothermal exponential hot halo of gas with a hydrogen density ($n_{\rm H}$) of $n_{\rm H}\sim3\times10^{-4}$~cm$^{-3}$ at $50$~kpc consistent with \citet{JBH2007}.
The gas is assumed to consist of small clouds in hydrostatic equilibrium with a hot halo.
The mean density of hydrogen around the dwarf galaxy is assumed to be $n_{\rm H}\sim10^{-3}$~cm$^{-3}$.
This density corresponds a filling factor of $\sim$$10$\% for the gas clouds at galactocentric distances of $150$~kpc.
Such gas would be easy to see in absorption through QSO sight lines but unable to be seen in emission.

In order for these small gas clouds to participate in the next starburst it must first fall back into the central gravitational potential and cool into a neutral cold phase.
For a low metallicity gas, $[$Fe$/$H$]=0.1$, at these densities the cooling timescale is $t_{\rm cool}\sim30$~Myr \citep{Sutherland1993}.
The cooling time will be slightly larger than the values given here due to the need to radiate energy gained from the collapse and the heating effect of the radiation fields.
The timescale of infall can be over an order of magnitude larger, $t_{\rm infall}\sim300$~Myr.
This infall timescale is comparable to the timescale of a typical burst of a few hundred megayears \citep{Lee2009,McQuinn2010b} and may influence how long the burst occurs for.
Both these timescales however, are much shorter than the orbital timescale of a few gigayear which determines how rapidly the Roche sphere changes.

This short timescale for cooling can only occur at apogalacticon.
At perigalacticon the gas will be extended along the orbital path according to Kepler's second law, lowering its density by a factor of $(1+\epsilon)/(1-\epsilon)$ and hence increasing its cooling time by a factor of $[(1+\epsilon)/(1-\epsilon)]^2$.
In addition, the gas experiences a much stronger radiation field which will counteract any cooling and reassociation taking place, leaving most gas ionized until apogalacticon.

Gas that has cooled and fell into the potential well of the dwarf may still be prevented from triggering another burst if it is kept ionized by the radiation field of the Galaxy.
As star formation requires optically thick gas to occur, all ionizing photons from the Galactic field are assumed to be absorbed, with the Galactic field dropping off as an inverse square law, with the (opacity corrected) photon flux given by \citep{JBH1999,JBH2002}
\begin{equation}
  \varphi = \frac{2.7\times10^8}{(R~{\rm kpc})^2}\frac{{\rm SFR}(t)}{{\rm SFR}(t_{\rm now})}~{\rm photons~cm}^{-2}~{\rm s}^{-1}.
\end{equation}

In addition to this Galactic radiation field, the dwarf will also experience the extragalactic UV background.
We use the time varying extragalactic UV field derived by \citet{Faucher-Giguere2009}, with the absorption of this radiation is done according to the prescription in \citep{Sternberg2002} with a constant density neutral core surrounded by an ionized medium of equal density.

\section{Results}

The timing of apogalacticons---and consequently bursts in this model---is determined by the orbital properties of the dwarf.
Two models are investigated, that of a simple Keplerian system described by equations (\ref{eq:rperi}) and (\ref{eq:rapo}) and one of a point mass dwarf orbiting inside a growing Einasto halo.
Both systems consist a dwarf with a point mass $m=1\times10^9$~$M_\odot$, a value consistent with the virial mass of a dwarf that formed at $z\sim10$ and possesses a dynamical mass similar to that observed in dwarf galaxies today within $300$~pc of the dwarf's center \citep{Strigari2008}.
Each dwarf has an initial gas mass of $1.5\times10^{8}$~$M_\odot$, giving the dwarf an approximately cosmological baryon to dark matter ratio.

Inside the Keplerian system the dwarfs orbit a host Galaxy of mass $M=5.3\times10^{11}$~$M_\odot$.
This mass is roughly half that of the virial mass of the Milky Way, but is consistent with the mass of the halo within $100$~kpc that reaches the same virial mass at the Milky Way derived from the RAVE survey \citep{Smith2007}.
These dwarf galaxies begin at apogalacticon $12$~Gyr ago, and experience a burst at every future apogalacticon.

For dwarfs inside a growing Einasto halo, a $z=0$ virial mass of $1.4\times10^{12}$~$M_\odot$ is assumed, consistent with the circular velocity from the RAVE survey.
The mass of this halo is calculated backward in time by assuming the median rate of growth for Milky Way size dark matter halos within the Millennium-II simulation \citep{Boylan-Kolchin2010}.
Dwarf galaxies inside the growing Einasto halo are assumed to be just past perigalacticon today and traced back in time $10$~Gyr.
The choice to calculate the orbit backwards from a beginning point---as opposed to the more realistic case of beginning at apogalacticon as in the Keplerian case---was taken to ensure a broad range of present day perigalacticons and eccentricities can be easily calculated as both of these quantities are not conserved in a growing halo.
As orbits changed substantially due to the Galaxy growth in mass between $8$--$10$~Gyr ago, orbits which have a perigalacticon during this time period, which was likely to be much larger than the perigalacticon today, were unlikely to have experienced any burst in this model with effected dwarfs behaving much more like isolated dwarfs.
The choice of tracing orbits back from perigalacticon was done to minimise the occurrence of this timing of perigalacticons with minimal differences otherwise. 
Orbits which still had a perigalacticon here were calculated from a point up to $10\%$ further along its orbit in order to see if most dwarfs of that perigalacticon and eccentricity combination would have experienced bursts.

\begin{figure}
  \begin{center}
    \includegraphics[width=0.5\textwidth]{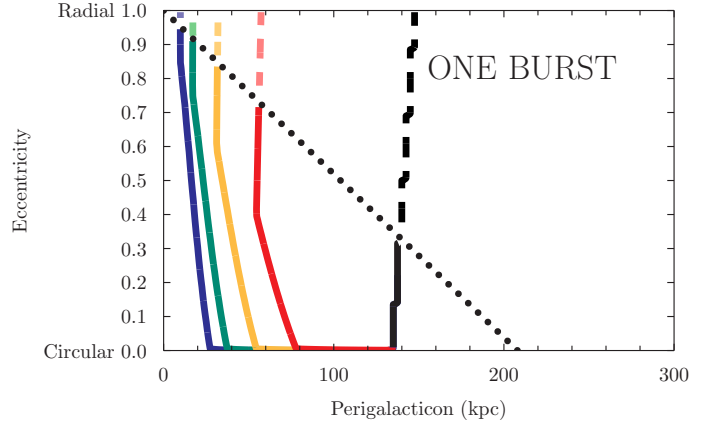}
  \end{center}
  \caption{The orbit properties (perigalacticon radius versus orbital eccentricity) of dwarf galaxies that retain sufficient gas to experience a burst of star formation post initial star formation in a Keplerian model with dwarf galaxy of mass $10^{9}$~$M_\odot$ and Galaxy point mass of $5.3\times10^{11}$~$M_\odot$.
The solid contours (left to right and blue to red in online version) correspond to gas masses of $1$, $3$, $10$ and $30\times10^6$~$M_\odot$ at the time of the first burst.
The dashed contours (light colors in online version) correspond to dwarfs that have yet to experience a burst of star formation but are expected to have sufficient gas to cause one in the future.
The dashed black line shows the upper limit of perigalacticons which experience bursts of any gas mass.
Beyond this, we consider dwarfs to be isolated, with a large Roche sphere that only slowly evolves.
Above the dotted line the orbit times exceed the age of the universe.
\label{fig:B1}}
\end{figure}

The amount of neutral gas available to fuel star formation as a function of perigalacticon and eccentricity is calculated from applying the time-varying Galactic and extragalactic radiation fields to the total mass of gas at any given apogalacticon given by equation (\ref{eq:Eq5}).
Many dwarfs are able to experience at least one burst subsequent to initial star formation, the mass available in the Keplerian system is shown in Figure \ref{fig:B1}, with dwarfs on approximately circular orbits far away experiencing the biggest bursts, a consequence of their large perigalacticon Roche sphere protecting much more gas from stripping.
Assuming that the halo mass of the Galaxy, the star formation rate in the Galaxy and the extragalactic UV background are constant into the future, the amount of gas available at a future burst is also able to be easily calculated. 

\begin{figure}
  \begin{center}
    \includegraphics[width=0.5\textwidth]{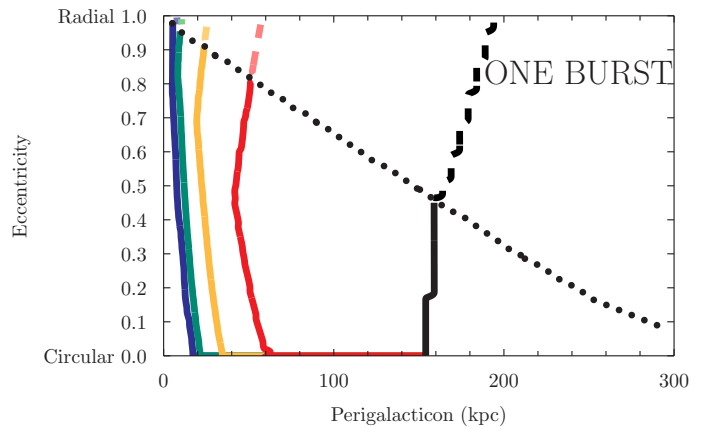}
  \end{center}
  \caption{The orbit properties (perigalacticon radius versus orbital eccentricity) of dwarf galaxies that retain sufficient gas to experience a burst of star formation post initial star formation in an Einasto halo.
The contours are detailed in Figure \ref{fig:B1}.
To the right of the black contour dwarfs have slowly evolving large Roche spheres and may be considered isolated.
Above the dotted line the orbit time of dwarfs which are at perigalacticon today is longer than $10$~Gyr.
\label{fig:BE1}}
\end{figure}

Similar contours are seen within the growing Einasto halo with the amount of gas available shown in Figure \ref{fig:BE1}.
As the Galaxy was less massive in the past a larger number of dwarfs of a given perigalacticon and eccentricity today were able to complete a full orbit and experience a burst of star formation.

\begin{figure}
  \begin{center}
    \includegraphics[width=0.5\textwidth]{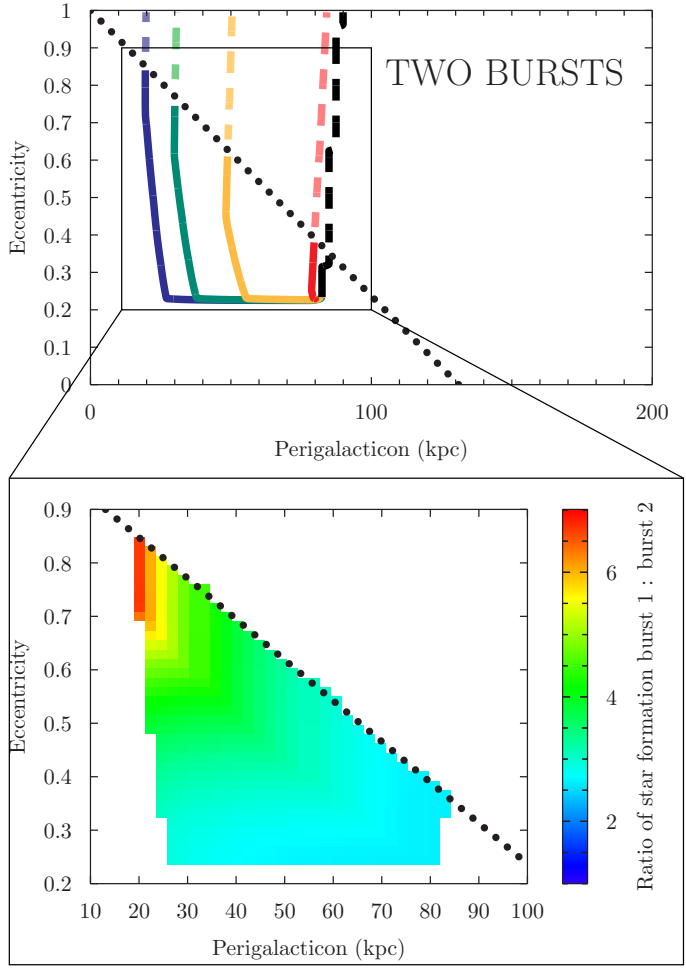}
   \end{center}
  \caption{The orbit properties (perigalacticon radius versus orbit eccentricity) that retain sufficient gas to experience a {\it second} burst of star formation in a Keplerian model.
The contours are detailed in Figure \ref{fig:B1}.
Above the dotted line orbit times exceed half the age of the universe.
The zoomed graph underneath shows the ratio of star formation that occurred in the first to the second burst in dwarfs that will have experienced two bursts today with at least $10^6$~$M_\odot$ of gas.
\label{fig:B2}}
\end{figure}

\begin{figure}
  \begin{center}
    \includegraphics[width=0.5\textwidth]{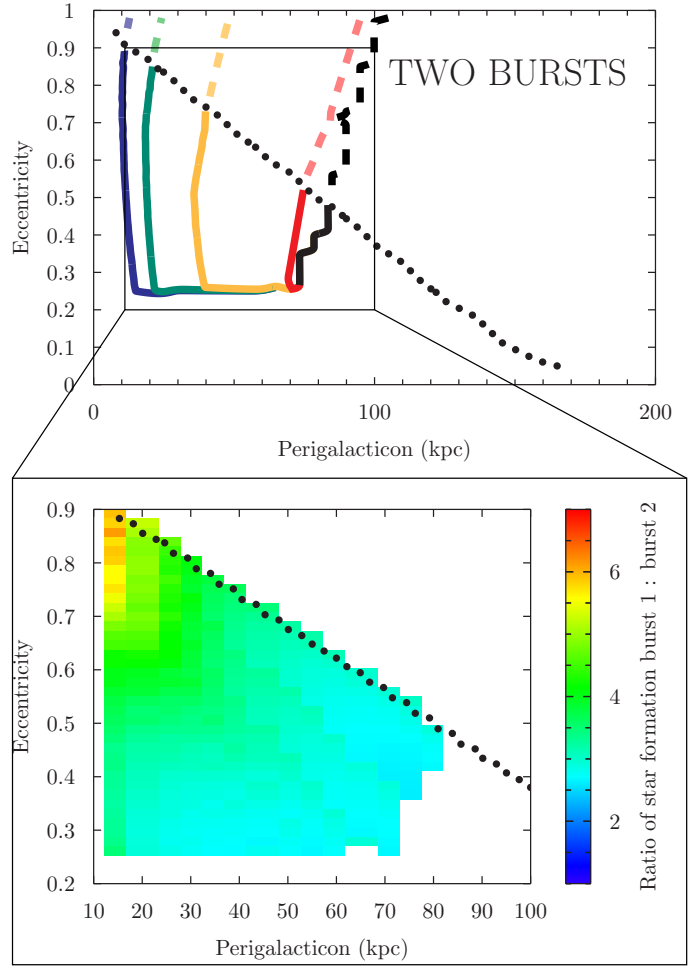}
   \end{center}
  \caption{The orbit properties (perigalacticon radius versus orbit eccentricity) that retain sufficient gas to experience a {\it second} burst of star formation in an Einasto model.
The contours are detailed in Figure \ref{fig:B1}.
Above the dotted line orbit times are long enough that dwarfs at perigalacticon today could not have completed two orbits in $10$~Gyr.
The zoomed graph underneath shows the ratio of star formation that occurred in the first to the second burst in dwarfs that will have experienced two bursts today with at least $10^6$~$M_\odot$ of gas.
\label{fig:BE2}}
\end{figure}

The allowed parameter space of dwarfs possessing large amounts of neutral gas ($M_{\rm gas}>10^6$~$M_\odot$) is greatly decreased for the second burst--the the third period of star formation.
In both Keplerian system dwarfs, shown in Figure \ref{fig:B2}, and Einasto system dwarfs, shown in Figure \ref{fig:BE2}, orbits at low perigalacticon are not able to hold onto sufficient amounts of gas for a second burst.
Similarly dwarfs with high perigalacticon in both systems have not had sufficient time to undergo multiple bursts.
These dwarfs may be able to undergo future bursts if given enough time, and assuming the ionization field is similar to that which exists today.

Very few regions exist which allow dwarfs to undergo a third burst in either system, with extremely few dwarfs ($<1\%$) having sufficient time to have completed three orbits while located at high enough perigalacticon to both shield sufficient amounts of gas and be located far enough from the Galaxy to protect dwarfs from the Galactic radiation field.

The star formation rate of a burst can be calculated by the amount of neutral hydrogen present within a dwarf, with the star formation rate in dwarfs being able to be approximated as a power law of the neutral gas mass ${\rm SFR}\propto M_{\HIsub}^{1.4}$ \citep{Kaisin2006}.
As the second burst has less gas, a consequence of equation (\ref{eq:Eq5}), this rate of star formation nearly always decreases.
Assuming that each burst is undertaken for an equal period of time, the ratio of the star formation in the first and second bursts can be calculated.
This ratio, also shown in Figures \ref{fig:B2} and \ref{fig:BE2}, is typically about three to one increasing as orbits become more radial $(\epsilon\rightarrow1)$.
This ratio always exceeds unity for dwarfs with over $10^6$~$M_\odot$ of H{\small I} available for star formation but can be less for close in, nearly circular orbits which have smaller amounts of gas available.
Dwarfs that have large amounts of gas ionized, that is dwarfs with a low apogalacticon occurring close at a time of heightened Galactic star formation, will not undergo a burst and may be able to hold onto enough gas for a future burst which is not accounted for in this model.

Assuming a common infall time of $10$--$12$~Gyr \citep[consistent with early infalling dwarfs][]{Rocha2011} and using orbit distribution results from {\it N}-body distributions \citep{Wetzel2011} we calculate the proportion of dwarfs that have experienced a first, second or third burst.

Around a Milky Way sized halo, up to $25\%$ of early infalling dwarfs in a Keplerian system should have experienced at least one burst with over $10^6$~$M_\odot$ of gas available.
A third of these dwarfs, $9\%$ of all early infalling dwarfs will have experienced two such bursts, while no dwarfs today are expected to have undergone three bursts of such magnitudes with any individual dwarf having only a $1\%$ chance of having completed three orbits each with a burst in the age of the Universe.
Within a more realistic growing Einasto halo, up to $40\%$ of early infall dwarfs will have experienced at least one burst \citep[this figure may be higher still due to the tendency for orbits to have higher perigalacticon inside extended halos][]{Wetzel2011}, half of these $17\%$ will have experienced two such bursts subsequent to initial star formation and $1.5\%$ three such bursts.

Late infalling dwarfs may have bursts of star formation that occurred when isolated from the hot halo environment \citep[e.g. Leo~I is likely to have had a burst of star formation at or before its recent infall][]{Rocha2011}.
Such dwarfs, if possessing gas when entering the hot halo, may only have undertaken this mechanism of bursts once if at all and would need to be considered separately from dwarfs that fell in while the Galaxy was young.

\section{The case of Carina and Fornax}

\begin{figure}
  \begin{center}
    \includegraphics[width=0.5\textwidth]{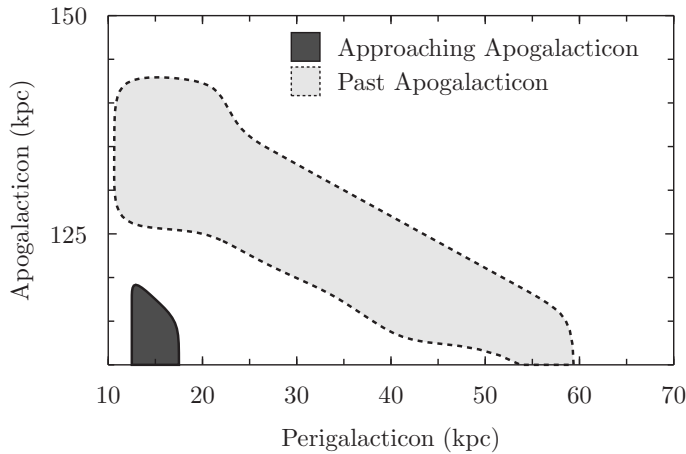}
  \end{center}
  \caption{The allowed orbits of Carina which produce two bursts containing over $10^{6}$~$M_\odot$ and have the first and third apogalacticon within $500$~Myr of the start of the bursts derived in \citet{Hurley-Keller1998}.
    The second apogalacticon is unlikely to have a burst due to the long period of star formation occurring at the first burst, which occurs over an entire orbit.
    Extending the allowed variation of bursts to $\pm1000$~Myr greatly increases the allowed parameter space of the orbits.
\label{fig:Carina}}
\end{figure}

The Carina dwarf spheroidal has three distinct periods of rapid star formation, separated by $\sim$$5$~Gyr.
Orbital restrictions from the proper motion of Carina \citep[a perigalacticon of $R_{\rm peri}=50\pm30$~kpc and apogalacticon of $R_{\rm apo}=110\pm30$~kpc][]{Lux2010,Pasetto2011} indicate that Carina will have undergone numerous apogalacticons since its infall $7$--$9$~Gyr ago \citep{Rocha2011}.
A number of orbits satisfy these orbital constraints while having apogalacticons, and therefore bursts in this model, occurring near the periods of star formation in Carina.
All these orbits have an apogalacticon in between the bursts of Carina, however, the length of Carina's first burst (second period of star formation) is comparable to the orbital period of Carina which will have prevented gas from cooling and fueling another burst immediately.
Carina's radial velocity is consistent with the dwarf both approaching and having just passed apogalacticon \citep{Piatek2003}.
This ambiguity leads to two distinct regions of allowed orbits in parameter space dependent upon where along its orbit Carina currently is.
These regions are bound by solid and dashed lines in Figure \ref{fig:Carina}.
If Carina is approaching apogalacticon today it is likely to be at a lower perigalacticon than its proper motion suggests and be almost at apogalacticon today.
Contrastingly, the orbits where Carina has just passed apogalacticon, the less likely scenario according to its radial velocity, are likely to be at either a larger perigalacticon, $R_{\rm peri}\sim50$~kpc or a higher apogalacticon, $R_{\rm apo}\sim130$~kpc compared to the case where it is approaching apogalacticon.
Only the case where Carina has recently passed apogalacticon is consistent with the proper motions, although low perigalacticon, high apogalacticon orbits are also allowed.

A greater constraint can be applied assuming that Carina's star formation rate would have consumed the gas within a Hubble time had it not been interrupted \citep[consistent with the star formation rates of gas rich dwarfs in the Local Group][]{Kaisin2006}.
To achieve the $\sim$$0.004$~$M_\odot$~yr$^{-1}$ rate calculated by \citet{Hurley-Keller1998} at this rate of consumption, Carina must have had $\sim$$5\times10^7$~$M_\odot$ of neutral hydrogen available.
Assuming that Carina has a dark matter mass of $10^9$~$M_\odot$ and began with $15\%$ of this mass in Hydrogen (consistent with the universal baryon to dark matter fraction) only the high pericentre orbits that have recently passed apogalacticon will have had sufficient gas to fuel Carinas star formation.
It must be noted however, that the gas consumption timescales differ by an order of magnitude amongst the Local Group dwarfs and hence Carina could conceivably be on any of the allowed orbits.

\begin{figure}
  \begin{center}
    \includegraphics[width=0.5\textwidth]{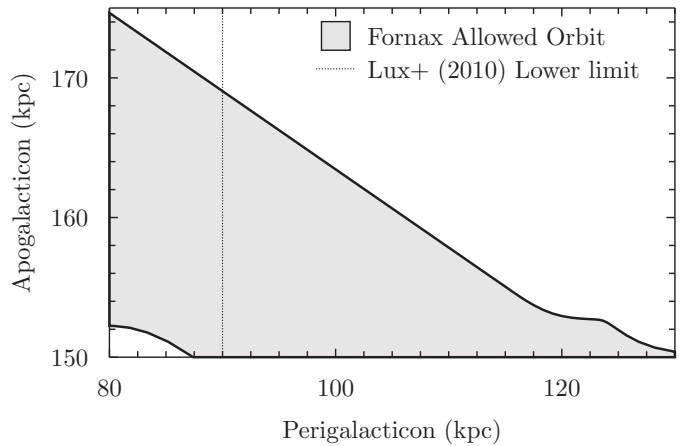}
  \end{center}
  \caption{The orbital range for which Fornax will have had only one burst with $>10^{6}$~$M_\odot$ of gas available between $3.5$ and $4.5$~Gyr ago.
    Combined with the lower perigalacticon limit from \citet{Lux2010} the allowed range is significantly reduced.
    \label{fig:Fornax}}
\end{figure}

This method can be extended to other dwarf spheroidals such as Fornax shown in Figure \ref{fig:Fornax}.
In order for Fornax's bursts to have occurred at the correct time its infall must have been towards the upper limit of $\sim8$~Gyr found by \citep{Rocha2011}.
However, this method clearly fails when applied to Ursa Minor dwarf which has one ancient period of star formation and no bursts post-infall \citep{Rocha2011}.
This is despite some allowed orbits of Ursa Minor predicted to have three bursts of star formation, this indicates that the protection of gas clouds by the dwarf may be overstated by this simple model.

\section{Conclusion}

We have presented a model for star formation bursts inside dwarf galaxies orbiting a host galaxy.
Under this model fall back of gas into the potential well of dwarf galaxies will occur at apogalacticon and allow episodic star formation for non-isolated dwarfs.
This model predicts that dwarfs with more than one burst today will be at a moderate perigalacticon from the host galaxy and of moderate eccentricity.
The second burst (the third period of star formation) is expected to be two to four times weaker than the first burst that follows initial star formation for most dwarfs.
This second burst may not have yet occurred in some dwarfs which, with a higher perigalacticon, may be expected to undergo another burst at a future point.
For dwarfs near apogalacticon and on the verge of experiencing their first or second burst of star formation we predict that low columns of diffuse ionized gas will be present in the vicinity of the dwarfs.
Such gas although much too faint to be seen in emission should have sufficient column to be detectable in absorption along QSO sight lines.
In particular this infalling gas should appear in recently infalling ($\lta5$~Gyr) dwarfs that are now near apogalacticon.
Leo~II is possibly the best candidate for this search.
\citet{Lepine2011} in calculating the proper motion suggest it is near either perigalacticon or apogalacticon, we would take the stronger view that due to its deficiency in gas \citep{Knapp1978} and the difficulty of stripping at such large radii \citep{Nichols2011} as well as the longer time spent at larger radii that it is near apogalacticon.
When combined with a recent infall \citep{Rocha2011} and the abundance of extragalactic sources nearby \citep{Lepine2011} Leo~II becomes a good candidate for  QSO sight lines picking up any warm gas infalling onto the dwarf.
Although Fornax is not near its expected apogalacticon \citep{Lux2010} the offset hydrogen feature although possibly Galactic gas \citep{Grcevich2009} and numerous QSO sight lines \citep{Tinney1999} would be worth investigating.

Dwarfs that have experienced no burst yet in this model, may be isolated enough to allow gas expelled from initial star formation to fall back into the dwarf triggering another burst \citep{Dong2003} and may have multiple bursts already, this process is unlikely to have happened in dwarfs that have already experienced one orbit of the Galaxy.

This model is able to place limits on the orbits of Carina and Fornax based upon the timing of their bursts after infall.
In particular, Carina is likely to have just passed apogalacticon, with a high perigalacticon and low apogalacticon.
For Carina to be explained by this model, the second apogalacticon will not have had a burst of star formation, with the dwarf still undergoing star formation from its first burst.
This extended period of star formation would have prevented a burst until the third apogalacticon.

As dwarfs that lose gas are likely to do so as gas clouds of not insignificant mass \citep{Mayer2006} there will be a large amount of variation as to how many clouds have managed to fall in by the beginning of star formation, and could result in a second burst being larger than the first unlike the results implied by this model.

\acknowledgements

D.C.L. is grateful to the University of Sydney for hosting
him during the preparation of this paper.
J.B.-H. is supported by a Federation Fellowship from the Australian Research Council.

\appendix
\section{Reaccretion with early internal heating}

\begin{figure}
  \begin{center}
    \includegraphics[width=0.5\textwidth]{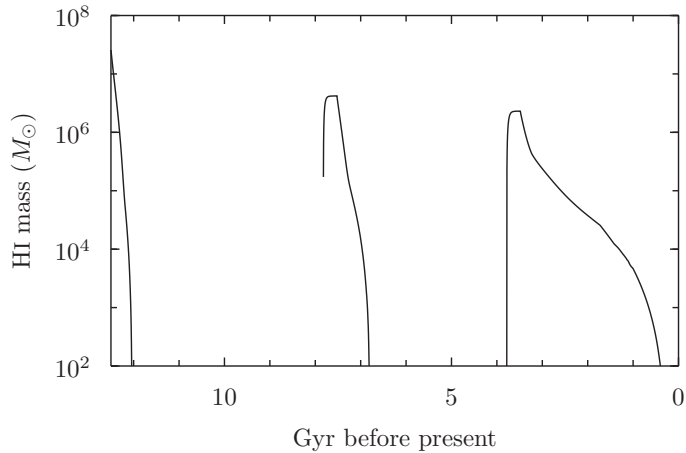}
  \end{center}
  \caption{Reaccretion added to \citet{Nichols2011} with cold gas added over $300$~Myr after apogalacticon. \label{fig:app}}
\end{figure}

The expulsion of gas from a dwarf is not automatic, and is helped greatly by internal heating assisted ram pressure stripping \citep{Nichols2011}.
Under instantaneous ram pressure stripping this gas, once lost, eventually ends up in the host.
However, gas once removed from the main body will experience less drag (arising from ram pressure) as the free stream velocity is reduced both by the dwarf galaxies shock and by the turbulent wake behind.
Even gas which is unbound at perigalacticon but sufficiently close may maintain enough momentum to later be reaccreted when far from the Galaxies center.
This gas can be modelled in the \citet{Nichols2011} case by the addition of cold gas after an apogalacticon (over a suitable period of time, dependent upon the freefall time) with the magnitude dictated by the above model.
A simple case of this is a Carina like orbit modelled from $z=5$ is shown in Figure \ref{fig:app}.
In this case, gas was added at the appropriate apogalacticons to simulate a Carina like burst.
The additional amount of gas added at the second period of star formation (the first burst) was quickly lost, this is due to the short period of time over which gas is added ($300$~Myr) and the stronger external radiation field present.
Although the first and second burst initially begin losing gas at the same rate, once mass is lost, the external radiation field---enhanced by the increased star formation of the Galaxy---has a larger effect on the first burst, resulting in a much more rapid gas loss than the second burst.

\bibliographystyle{apj}

\end{document}